\documentclass[a4paper,10pt,twoside]{cpc-hepnp}
\usepackage{CJK,upgreek,fancyhdr}
\usepackage{graphicx}
\usepackage{booktabs}
\usepackage{amssymb,bm,mathrsfs,bbm,amscd}
\usepackage[tbtags]{amsmath}
\usepackage{lastpage}
\usepackage{multirow}
\usepackage{color}
\usepackage{float}
\usepackage{lineno}
\usepackage{appendix}

\begin{document}
\begin{CJK*}{GB}{gbsn}

\fancyhead[c]{\small Chinese Physics C~~~Vol. xx, No. x (2020) xxxxxx}
\fancyfoot[C]{\small xxxxxx-\thepage}

%\footnotetext[0]{Received 31 June 2015}

\title{The Possible Equation Of State Of Dark Matter in Low Surface Brightness Galaxies
\thanks{supported by the National Natural Science Foundation
(NSF) of China (No. 11973081, 11573062, 11403092, 11390374, 11521303),
the YIPACAS Foundation (No. 2012048), the Chinese Academy of Sciences
(CAS, KJZD-EW-M06-01), the NSF of Yunnan Province (No. 2019FB006) and
the Youth Project of Western Light of CAS.}
}

\author{%
      Xiaobo Gong (¹¨Ð¡²¨)$^{1,2,3,4}$\email{gxbo@ynao.ac.cn}%
\quad Meirong Tang (ÌÆÃÀÈÙ)$^{1,2,3,4}$
\quad Zhaoyi Xu (ÐìÕ×Òâ)$^{5}$\email{xuzy@ihep.ac.cn}
}

\maketitle

\address{%
$^1$ Yunnan Observatories, Chinese Academy of Sciences, 396 Yangfangwang, Guandu District, Kunming, 650216, P. R. China\\
$^2$ Key Laboratory for the Structure and Evolution of Celestial Objects, Chinese Academy of Sciences, 396 Yangfangwang, Guandu District, Kunming, 650216, P. R. China\\
$^3$ Center for Astronomical Mega-Science, Chinese Academy of Sciences, 20A Datun Road, Chaoyang District, Beijing, 100012, P. R. China\\
$^4$ University of Chinese Academy of Sciences, Beijing, 100049, P. R. China \\
$^5$ Key Laboratory of Particle Astrophysics, Institute of High Energy Physics, Chinese Academy of Sciences, Beijing 100049, China
}

\begin{abstract}
The observed rotation curves of low surface brightness (LSB) galaxies play an essential role in studying dark matter, and indicate that there exists a central constant density dark matter core. However, the cosmological N-body simulations of cold dark matter predict an inner cusped halo with a power-law mass density distribution, and can't reproduce a central constant-density core. This phenomenon is called cusp-core problem. When dark matter is quiescent and satisfies the condition for hydrostatic equilibrium, using the equation of state can get the density profile in the static and spherically symmetric space-time. To solve the cusp-core problem, we  assume that the equation of state is independent of the scaling transformation. Its lower order approximation for this type of equation of state can naturally lead to a special case, i.e. $p=\zeta\rho+2\epsilon V_{rot}^{2}\rho$, where $p$ and $\rho$ are the pressure and density, $V_{rot}$ is the rotation velocity of galaxy, $\zeta$ and $ \epsilon$ are positive constants. It can obtain a density profile that is similar to the pseudo-isothermal halo model when $\epsilon$ is around $0.15$. To get a more widely used model, let the equation of state include the polytropic model, i.e. $p= \frac{\zeta}{\rho_{0}^{s}}\rho^{1+s}+ 2\epsilon V_{rot}^{2}\rho$, we can get other kinds of density profiles, such as the profile that is nearly same with the Burkert profile, where $s$ and $\rho_{0}$ are positive constants.
\end{abstract}

\begin{keyword}
galaxy: dark matter--galaxy: rotation curves
\end{keyword}

\begin{pacs}
xxxxxx
\end{pacs}

\footnotetext[0]{\hspace*{-3mm}\raisebox{0.3ex}{$\scriptstyle\copyright$}2019
Chinese Physical Society and the Institute of High Energy Physics
of the Chinese Academy of Sciences and the Institute
of Modern Physics of the Chinese Academy of Sciences and IOP Publishing Ltd}%

%\begin{multicols}{2}

\section{Introduction}

The dark matter(DM) is an unsolved puzzle in cosmology and particle physics, and it probably consists of particles that are weakly interacting. Though many astronomical observations, like Cosmic Microwave Background and baryon acoustic oscillations, approximately dark matter make up $23\%$ of today's Universe. The cosmological model based on cold dark matter in reproducing the large-scale structure of the Universe is quite well and get great success\cite{2006PhRvD..74l3507T,2008MNRAS.391.1685S,2009ApJS..180..225H,2011ApJS..192...18K,2010MNRAS.404...60R,2009ApJ...699..539R,2010A&A...516A..63S}. The most popular candidate for cold dark matter is the weakly interacting massive particles (WIMPs), which are particles with negligible self-interactions, they are stable  and collisionless. Their particle masses estimated are in the range 10 GeV $\sim$ 1 TeV.

 However, a plenty of serious challenges to the cold dark matter model  have emerged on the small scale, such as  the scale of individual galaxies and their central core\citep{2008MNRAS.391.1685S}. For example, in the cold dark matter model, halos can be characterized by a power-law mass density distribution with a steep power index at the central core, which is in contrary to observation in small scale, such as the observed rotation curves of low surface brightness (LSB) galaxies. It shows that a central constant-density dark matter core exists\citep{2001ApJ...552L..23D} in LSB galaxies, which consist of a very small proportion of ordinary baryonic matter so that their stellar populations make only a relatively small contribution to the observed rotation curves. The rotation curves of galaxies are important observational tools to detect  their gravitational potential. Since the luminous component can not be able to fit the whole rotation curve of galaxies, it needs to  add a contribution by dark matter or the other ideas, such as modified Newtonian dynamics (MOND)\cite{1983ApJ...270..365M,1983ApJ...270..371M}. When we use dark matter  to explain the observed rotation curves of galaxies, there exist a lot of empirical spherical dark matter halo profiles\cite{2017MNRAS.465.4703K}, such as the Navarro-Frenk-White (NFW) profile\cite{1997ApJ...490..493N}, Burkert profile\citep{1995ApJ...447L..25B} and pseudo-isothermal profile\cite{1991MNRAS.249..523B}. The Burkert profile and pseudo-isothermal profile are often used for LSB galaxies, they can perfectly  fit the observed rotation curves of LSB galaxies\cite{2006ApJS..165..461K,2019MNRAS.490.5451D,2020arXiv200503520D}. They both indicate there exist a central constant density. However, making use of the cold dark matter model, numerical N-body simulations  can't reproduce a central constant-density dark matter core\citep{1996ApJ...462..563N}. This phenomenon is called cusp-core problem.

Furthermore, LSB galaxies are valuable laboratories for the indirect detection of DM\cite{2021MNRAS.501.4238B}. They are late-type disc galaxies with a face-on central surface brightness fainter than that of the night sky. It is very difficult to  detect them due to their very low surface brightness. LSB galaxies are the most numerous objects in the Universe, and possibly
contribute  $\gtrsim 50\%$ to the galaxy population\cite{2020arXiv200503520D,2006A&A...458..341T}. They are generally isolated systems\cite{2019ApJ...879L..12K}. The pair annihilation of WIMPs can product high energy $\gamma$-rays\cite{{1985NuPhB.253..375S}}. Star formation, accreting supermassive black holes or active galactic nuclei (AGN) also can lead to $\gamma$-ray emission. LSB galaxies generally have very low star formation rates and are characterized by diffuse, metal poor and very low surface density exponential stellar discs. They are slowly evolving galaxies. AGNs are rarely discovered in LSB galaxies\cite{2013JApA...34...19D}.  Moreover, LSB galaxies are very rich in neutral hydrogen (${\rm H_{I}}$) gas. On average, the masses of ${\rm H_{I}}$ gas in their gas discs range  from $10^{8}$ to $10^{10} M_{\odot}$ \cite{2019ApJS..242...11L}. The extended gas discs  can extend outward to  about 2-3 times their stellar disks\cite{1996MNRAS.283...18D,2017MNRAS.464.2741M}. This enables us to measure their rotation curves out to large radii by using the ${\rm H_{I}}$ data. Their rotation curves also can be derived from the other emission lines such as  ${\rm H_{\alpha}}$ line\citep{2006ApJS..165..461K, 2008ApJ...676..920K,2010ApJ...710L.161K}. Rotation curve studies indicate LSB galaxies are strongly dominated by DM and have simple dynamical structure. 

 There exist many other famous DM models to explain the cusp-core problem, such as self-interacting dark matter\citep{2013MNRAS.430...81R} and warm dark matter model(WDM)\citep{2001ApJ...556...93B,2012MNRAS.424.1105M}. Two of the most canonical candidates for WDM are the sterile neutrino\citep{2019PrPNP.104....1B} and the gravitino\citep{2007JHEP...03..037B}. The dark matter particles are usually classified by their velocity dispersion given in terms of three broad categories: hot (HDM), warm (WDM) and cold (CDM) dark matter. In principle HDM is relativistic at all cosmological relevant scales. When a particle's momentum is  equal to or less than its mass, it becomes non-relativistic. The warm dark matter has a bigger velocity  than cold dark matter because of their mass. The typical mass of WDM  particle is around 1 KeV.  On subgalactic and galactic scales, their non-zero thermal velocities have a strong suppression effect on the steep dark matter power spectrum\citep{2012MNRAS.424.1105M}.

 In this paper,  we want to study the equation of state of dark matter to explain cusp-core problem.  Based on the nature of warm dark matter, we think dark matter has a nonzero random motion in the density core. However, when dark matter is a perfect fluid, the observations show that the velocity of this random motion is far less than the speed of light. In reality, this constant random motion probably vanishes at large radii, so this term  can be replaced by the polytropic model. Taking into account the properties of  WIMPs, we can assume that random motion of dark matter particles is positively correlated with their rotational motions at large scale. Moreover, because many halo density profiles, like NFW profile or pseudo-isothermal profile, can be used to fit the density profile of the galaxies from small sizes to large sizes effectively, we can assume that the equation of state is independent of the scaling transformation. In the paper, we will show that its lower order approximation for this type of equation of state can naturally yield that random motion of dark matter include a constant term and a term that is proportional to the particle rotational motions. In this simple phenomenological model, we can see the dark matter halo profile is in agreement with the observations. It can give a solution for the cusp-core problem.

 The outline of this paper is as follows. In Section 2, we present three special cases for the equation of state of dark matter and obtain the mass density profiles of dark matter halos.  We obtain an approximate analysis to describe the interactive effect between dark matter and black hole in subsection 2.1. In subsection 2.2, we study our above phenomenological model and compare it with the observations. In subsection 2.3, to let the constant random motions  vanish at large radii, this term  is replaced by the polytropic model.  We then study this new model. In Section 3, we present the conclusions in our work.

\section{The model}
\label{sect:model}

The most general static and spherically symmetric  metric  takes the following form
\begin{equation}
\begin{aligned}
ds^{2}=e^{A}dt^{2}-e^{B}dr^{2}-r^{2}(d\theta^{2}+\sin^{2}{\theta}d\phi^{2}),
 \end{aligned}
\end{equation}
where $A$ and $B$ are function of $r$. For conventions, the gravitational constant and the speed of light are set equal to 1.
Using the above metric, the Einstein gravitational field equation can be expressed as follows.
\begin{equation}
\begin{aligned}
&-e^{-B}(\frac{1}{r^{2}}-\frac{B_{r}^{\prime}}{r})+\frac{1}{r^{2}}=8\pi T_{t}^{t}, \\
&-e^{-B}(\frac{1}{r^{2}}+\frac{A_{r}^{\prime}}{r})+\frac{1}{r^{2}}=8\pi T_{r}^{r}, \\
&\frac{-e^{-B}}{2}[A_{rr}^{\prime\prime}+\frac{(A_{r}^{\prime})^{2}}{2}+\frac{A_{r}^{\prime}-B_{r}^{\prime}}{r}-\frac{A_{r}^{\prime}B_{r}^{\prime}}{2}]=8\pi T_{\theta}^{\theta}=8\pi T_{\phi}^{\phi}. \\
\label{eq:gravity1}
 \end{aligned}
\end{equation}
Through analyzing the stable circular orbits of a test particle, the rotation velocity $V_{rot}$ of a test particle can be expressed in the form\citep{2003astro.ph..3594M,2010PhLB..694...10R},
\begin{equation}
\begin{aligned}
V_{rot}^{2}=\frac{1}{2}rA_{r}^{\prime}.
\label{eq:gravity2}
 \end{aligned}
\end{equation}
If the pure dark matter is an isotropic perfect fluid, its energy momentum tensor  takes the form $T^{\mu}_{\nu}={\rm diag}[\rho, -p, -p, -p]$
for the spherically symmetric case.  Let us set $F=e^{-B}-1$, $N=A_{x}^{\prime}$ and $x=\ln(r)$.
Then Equation~(\ref{eq:gravity1}) become
%\begin{equation}
\begin{align}
&F+F_{x}^{\prime}+8\pi r^{2}\rho=0, \label{eq:gsim1}\\
&(1+F)N+F=8\pi  r^{2}p, \label{eq:gsim2}\\
&(1+F)\left[N_{x}^{\prime}+\frac{1}{2}N^{2}\right]+\frac{1}{2}F_{x}^{\prime}N+F_{x}^{\prime}=16\pi r^{2} p.\label{eq:gsim3}\\
 \end{align}
%\end{equation}
It yields
\begin{equation}
\begin{aligned}
\frac{dp}{dx}=-\frac{p+\rho}{2}N.
\end{aligned}
\end{equation}
This equation  can lead to $\frac{dp}{dr}=-\rho g$ in non-relativistic approximation, where $g$ is the gravity acceleration. Then in relativistic case, we have the following equation
\begin{equation}
\begin{aligned}
N_{x}^{\prime}+\left[\frac{F_{x}^{\prime}}{2(1+F)}-2\right]N+\frac{N^{2}}{2}+\frac{F_{x}^{\prime}-2F}{1+F}=0.
\label{eq:gravity_aa1}
 \end{aligned}
\end{equation}
To solve this equation, we need the equation of state(EOS).
Because combining Equation~(\ref{eq:gsim1}) and Equation~(\ref{eq:gsim2})  yields the following expression
\begin{equation}
\begin{aligned}
\frac{p}{\rho}=-\frac{(1+F)N+F}{F+F_{x}^{\prime}},
\label{eq:gravity_a1}
 \end{aligned}
\end{equation}
and assume that pressure $p$ is only the function of density $\rho$,  this lead to $N$ is a function of  $x, F_{x}^{\prime}$ and $F$, and mark it as $N=G(x,F,F_{x}^{\prime})$.
If $F(r)$ is a solution and $\lambda$ is a positive constant, then $F(\lambda r)$ is also its solution. This assumption leads to the equation $N=G$
doesn't obviously contain $x$. It is an autonomous equation. However, this also leads to the pressure $p$ is proportional to the density $\rho$.
To take into account more possible EOS, we just assume that $N$ is a function of  $F_{x}^{\prime}$ and $F$. The rotation velocity is far less than the speed of light so that $F$ is very small. When $F$ and $F_{x}^{\prime}$ are very small,
we can use the Taylor expansion of  equation $G(F,F_{x}^{\prime})$ to stand for it.
Because the equation (\ref{eq:gravity_aa1}) can be solved by iteration,
 \begin{equation}
\begin{aligned}
&N_{0}=-F \\
&N_{1}=-F(1+\frac{3F_{x}^{\prime}-5F-F^{2}}{4+4F-F_{x}^{\prime}})\approx -F(1+\frac{3}{4}F_{x}^{\prime}-\frac{5}{4}F)\\
&N_{2}\approx -F(1-\frac{1}{2}F_{x}^{\prime}-\frac{5}{4}F+\frac{3}{8}F_{xx}^{\prime\prime})-\frac{3}{8}(F_{x}^{\prime})^{2} \\
& ...
\label{eq:gravity_a2}
 \end{aligned}
\end{equation}
taking  into account the above approximate iteration form of variable $N$, we consider two cases of the Taylor expansion of  equation $N=G(F,F_{x}^{\prime})$ to solve the Equation (\ref{eq:gravity_aa1}) in the following subsection 2.1 and 2.2.  We then take  into account an extended  EOS of dark matter in subsection 2.3. This EOS is not an autonomous equation.

\subsection{Case I}
\label{sect:model1}

Using the first iteration formula $N_{1}$, we can assume
\begin{equation}
\begin{aligned}
N=-F+(\gamma F+\epsilon F_{x}^{\prime})F,
\label{eq:gravity_b1}
 \end{aligned}
\end{equation}
where $\gamma$ and $\epsilon$ are constant.  When $|F|\ll 1$ and $|F_{x}^{\prime}|\ll 1$, using Equation~(\ref{eq:gravity2}), Equation~(\ref{eq:gsim1}) and Equation~(\ref{eq:gsim2}),
this assumption leads to the following equation of state (EOS)
\begin{equation}
\begin{aligned}
p=\frac{(1+F)N+F}{8\pi r^{2}}\approx\frac{\epsilon(F_{x}^{\prime}+F)F+(\gamma-\epsilon-1)F^{2}}{8\pi r^{2}}\approx 2\epsilon V_{rot}^{2}\rho+\frac{\gamma-\epsilon-1}{2\pi}(\frac{V_{rot}^{2}}{r})^{2}.
\end{aligned}
\end{equation}
When $|F|\ll 1$ and $|F_{x}^{\prime}|\ll 1$, and setting $M=\ln(-F)$,  the Equation (\ref{eq:gravity_aa1}) and Equation (\ref{eq:gravity_b1}) can be approximated by
\begin{equation}
\begin{aligned}
2\epsilon U_{x}^{\prime}+(4\gamma-4\epsilon-3)U+4\epsilon U^{2}+5-4\gamma=0,
\label{eq:gravity_b2}
 \end{aligned}
\end{equation}
where $U=M_{x}^{\prime}$. This equation is one kind of Riccati Equation.
when $(4\gamma+4\epsilon-3)^{2}-32\epsilon >0$, one of the solutions is
\begin{equation}
\begin{aligned}
F=-\frac{b}{r^{\widetilde{\alpha}}}\sqrt{ 1+(\frac{r}{r_{0}})^{\widetilde{\beta}}}, ~( \widetilde{\beta} > 0)
\label{eq:gravity_b3}
 \end{aligned}
\end{equation}
where $\widetilde{\alpha}=\frac{4\gamma-4\epsilon-3}{8\epsilon}+\frac{\widetilde{\beta}}{4}$, $\widetilde{\beta}=\frac{\sqrt{(4\gamma+4\epsilon-3)^{2}-32\epsilon}}{2|\epsilon|}$, $b$ and the core radius $r_{0}$ are positive constant.
Then the density $\rho$ is
\begin{equation}
\begin{aligned}
\rho=-\frac{F}{8\pi r^{2}}\left[1-\widetilde{\alpha}
+\frac{\widetilde{\beta}(\frac{r}{r_{0}})^{\widetilde{\beta}}}{2+2(\frac{r}{r_{0}})^{\widetilde{\beta}}}\right].
\end{aligned}
\end{equation}
If density is characterized by a power-law distribution $\rho \sim r^{\alpha}$, using Equation~(\ref{eq:gsim1}), $\alpha$ can be written as
\begin{equation}
\begin{aligned}
\alpha=\frac{r\rho_{r}^{\prime}}{\rho}=\frac{F_{x}^{\prime}+F_{xx}^{\prime\prime}}{F+F_{x}^{\prime}}-2.
\end{aligned}
\end{equation}
The absolute value of the slope $\alpha$ should be higher in the outer region.
Considering the fact that dark matter  density  $\rho$ has a lower value and its absolute value of slope $\alpha$ is higher in the outer region, $\widetilde{\alpha}$ should be equal to 1.
This lead to $\gamma=1+\epsilon$, then
it is easy to see that $p $ is proportional to the term $V_{rot}^{2}\rho$, i.e. $p\approx 2\epsilon V_{rot}^{2}\rho$ (i.e. $\frac{(1+F)N+F}{8\pi r^{2}}=-\epsilon N\frac{F_{x}^{\prime}+F}{8\pi r^{2}}$), and the variable $N$ is given by
\begin{equation}
\begin{aligned}
N=- \frac{F}{1+F+\epsilon F+\epsilon F_{x}^{\prime}}.
\label{eq:gravity8}
\end{aligned}
\end{equation}
Thus, using Equation (\ref{eq:gravity_aa1}) can lead to the following equation
\begin{equation}
\begin{aligned}
2\epsilon F_{xx}^{\prime\prime}+F_{x}^{\prime}(1-\epsilon+\frac{\epsilon-6\epsilon^{2}F}{1+F})
+\frac{\epsilon (F_{x}^{\prime})^{2}}{F}(1+\frac{1+2\epsilon F_{x}^{\prime}}{1+F})
+(1-4\epsilon)F-\frac{4\epsilon^{2}F^{2}}{1+F}=0,
\label{eq:gravity3}
 \end{aligned}
\end{equation}
and Equation (\ref{eq:gravity_b3}) is reduced into the form
\begin{equation}
\begin{aligned}
F=-\frac{b}{r}\sqrt{ 1+(\frac{r}{r_{0}})^{\frac{8\epsilon-1}{2\epsilon}}}.
\label{eq:gravity6}
 \end{aligned}
\end{equation}
When the core radius $r_{0}\rightarrow\infty$, this solution can become the vacuum Schwarzschild solution and the parameter $b$ is the Schwarzschild radius in this case. Then the energy density $\rho$ can be approximated as
\begin{equation}
\begin{aligned}
\rho=\frac{(8\epsilon-1)b}{32\epsilon\pi r_{0}^{3}}\frac{(\frac{r}{r_{0}})^{\frac{2\epsilon-1}{2\epsilon}}}{\sqrt{ 1+(\frac{r}{r_{0}})^{\frac{8\epsilon-1}{2\epsilon}}}}.
\label{eq:gravity7}
\end{aligned}
\end{equation}
This profile  is one of the Zhao halos profile \citep{1996MNRAS.278..488Z}, which can acquire both the form of a cusped or a cored profile with three free parameters ($\overline{\alpha}, \overline{\beta}, \overline{\gamma}$):
\begin{equation}
\begin{aligned}
\rho=\frac{\rho_{0}}{(\frac{r}{r_{0}})^{\overline{\gamma}}( 1+(\frac{r}{r_{0}})^{\overline{\alpha}})^{\frac{\overline{\beta}+\overline{\gamma}}{\overline{\alpha}}}}.
\label{eq:gravity7a3}
\end{aligned}
\end{equation}
When $r_{0}$ is very large, there exists a transition zone between density core and outer region when density is characterized by a power-law distribution $\rho \sim r^{\alpha}$ for the profile in the Equation (\ref{eq:gravity7}). In this transition zone, the slope is changing from $\alpha=\frac{2\epsilon-1}{2\epsilon}$ to $\alpha=-\frac{1+4\epsilon}{4\epsilon}$. For example, when $\epsilon=0.5$, the density distribution is dominated by a central constant-density core, and is dominated by an outer power-law density distribution $\rho \sim r^{-1.5}$. Because in the outer region dark matter has lower density, the interaction force of the dark matter becomes smaller, then variable $\epsilon$ may be smaller at larger radii, we probably get a  steeper outer power-law density distribution  in this more realistic case, like the pseudo-isothermal profile $\frac{\rho_{0}}{1+(r/r_{0})^{2}}$. Unfortunately, there exists a serious problem. Because $b$ is the Schwarzschild radius, so the mass of dark matter within the core radius is only $\sqrt{2}-1$ times the black hole mass. This is not consistent with the observation.  The observations show that the mass of the central black hole is far less than that of the dark matter halo for many types of galaxies\citep{2016MNRAS.455.3148S,2006MNRAS.373..700B,2021arXiv210510508M}.

\begin{figure}[H]
  \centering
  \includegraphics[angle=0,height=8.0cm,width=10.0cm,bbllx=110pt,bblly=219pt,bburx=496pt,bbury=556pt]{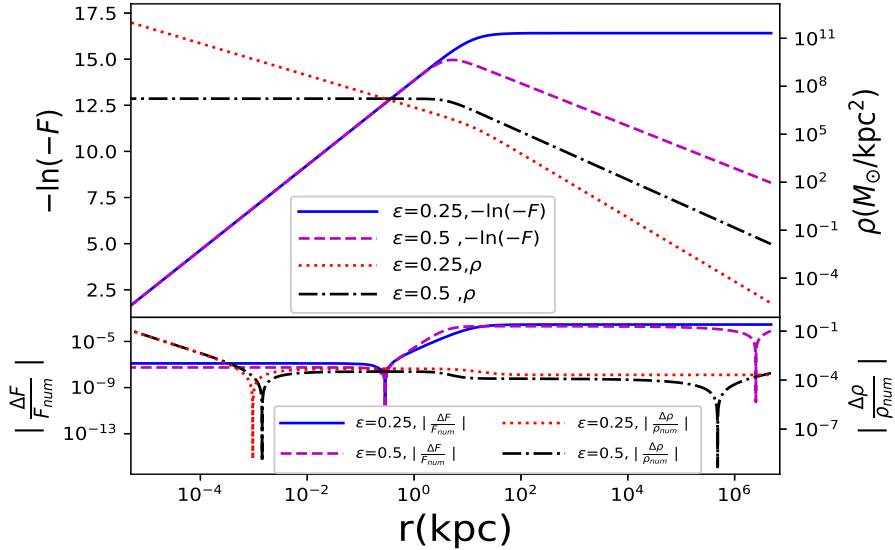}\\
    \caption{\small  Comparison results between the approximate analysis and the numerical solutions. \textbf{Top}: the analytical  solutions of $-\ln(-F)$ and $\rho$. The solid line and the dashed line stand for $-\ln(-F)$ when  $\epsilon$=0.25 and 0.5, respectively. The dotted line  and the dash-dotted line represent $\rho$ when $\epsilon$=0.25 and 0.5, respectively. \textbf{Bottom}: The relative errors ($\mid\frac{\Delta F}{F_{num}}\mid=\mid\frac{ F_{num}-F_{app}}{F_{num}}\mid$ and $\mid\frac{\Delta \rho}{\rho_{num}}\mid=\mid\frac{ \rho_{num}-\rho_{app}}{\rho_{num}}\mid$) for different $\epsilon$. The subscripts $num$ and $app$ refer to quantities relate to the numerical solutions and the the approximate analysis, respectively. The solid line and the dashed line correspond to $\mid\frac{\Delta F}{F_{num}}\mid$ when $\epsilon$=0.25 and 0.5, respectively. The dotted line and the dash-dotted line give $\mid\frac{\Delta \rho}{\rho_{num}}\mid$ when $\epsilon$=0.25 and 0.5, respectively.
     }
  \label{Fig:dis1}
\end{figure}

To test the approximate analysis in the Equation (\ref{eq:gravity6}), we use the odeint Python routine in the SciPy library to solve the Equation (\ref{eq:gravity3}). Because the energy density $\rho$ is not easy to be solved by Equation (\ref{eq:gravity1}),
using Equation (\ref{eq:gravity1}) and Equation (\ref{eq:gravity3}),
the energy density $\rho$ can be rewritten as
\begin{equation}
\begin{aligned}
\ln(8\pi r^{2}\rho)_{x}^{\prime}=\frac{F_{xx}^{\prime\prime}+F_{x}^{\prime}}{F_{x}^{\prime}+F}
=-\frac{1}{2\epsilon}\Bigg[1-4\epsilon+\frac{\epsilon F_{x}^{\prime}}{F}+\frac{\epsilon F_{x}^{\prime}}{F(1+F)}-\frac{2\epsilon^{2}F_{x}^{\prime}}{1+F}
+\frac{2\epsilon^{2}(F_{x}^{\prime})^{2}}{F(1+F)}-\frac{4\epsilon^{2}F}{1+F}\Bigg].
\label{eq:gravity4}
 \end{aligned}
\end{equation}
Then, using the Equation (\ref{eq:gravity4}) obtains the density $\rho$.
In Fig. \ref{Fig:dis1}, we compared our approximated analytical solutions with the full numerical solutions for $F$ and $\rho$. It is obvious to see that their difference is tiny, i.e. the relative differences are generally below $10^{-4}$, mostly stay at $10^{-4}-10^{-5}$ level.   The relative errors of density $\rho$ become larger and can get $10^{-1}$ at the inner region near the black hole.  The analytical solution can give the form of density profile directly.

\subsection{Case II}
\label{sect:model3}
In the previous case, there does not exist the solution that satisfies the boundary  conditions
\begin{equation}
\begin{aligned}
&F=0, ~~\alpha\approx 0 ~~~{\rm at}~~~ r=0 \\
&\alpha\approx -2 ~~~{\rm at}~~~ r=\infty.
\end{aligned}
\end{equation}
 Assume that
\begin{equation}
\begin{aligned}
N=-F-\zeta(F+F_{x}^{\prime})+(1+\epsilon )F^{2}+\epsilon F_{x}^{\prime}F,
\label{eq:gravity_bbb1}
 \end{aligned}
\end{equation}
where $\zeta$ is a very small positive constant\citep{2010PhLB..694...10R,2016PhLB..753..140P}.
When $|F|\ll 1$ and $|F_{x}^{\prime}|\ll 1$, using Equation~(\ref{eq:gravity2}), Equation~(\ref{eq:gsim1}) and Equation~(\ref{eq:gsim2}),
this assumption leads to the following equation of state (EOS)
\begin{equation}
\begin{aligned}
p=\frac{(1+F)N+F}{8\pi r^{2}}\approx\frac{-\zeta(F+F_{x}^{\prime})+\epsilon(F_{x}^{\prime}+F)F}{8\pi r^{2}}\approx \zeta\rho+ 2\epsilon V_{rot}^{2}\rho.
\end{aligned}
\end{equation}
When $|F|\ll 1$ and $|F_{x}^{\prime}|\ll 1$, the Equation (\ref{eq:gravity_aa1}) and Equation(\ref{eq:gravity_bbb1}) can lead to the following approximate equation:
\begin{equation}
\begin{aligned}
(2\epsilon H-1)H_{xx}^{\prime\prime}+(1+H)H_{x}^{\prime}+2\epsilon(H_{x}^{\prime})^{2}
+2H+(1-4\epsilon)H^{2}=0.
\label{eq:app2}
\end{aligned}
\end{equation}
Where $H=\frac{F}{2\zeta}$.
When $|F| \ll \zeta$, this case leads to the following approximate equation
\begin{equation}
\begin{aligned}
F_{xx}^{\prime\prime}-F_{x}^{\prime}-2F=0.
\end{aligned}
\end{equation}
Then its solution is
\begin{equation}
\begin{aligned}
F=-\frac{b}{r}-(\frac{r}{r_{0}})^{2},
\label{eq:gravity_blak_hole}
\end{aligned}
\end{equation}
where $b$ and  $r_{0}$ are constant. This solution can include a black hole or a constant-density core. A black hole and a constant-density dark matter core can  hold simultaneously in one halo. When $b=0$, the black hole will not exist. It is a good approximate solution near the halo center.  When $\zeta \ll |F| \ll 1, ~\epsilon \geq\frac{1}{4}$
and $F_{x}^{\prime}=\chi F$, using the Equation (\ref{eq:app2}), we obtain
\begin{equation}
\begin{aligned}
\chi=\frac{4\epsilon-1}{4\epsilon} ~~~{\rm or}~~~ \alpha= -\frac{4\epsilon+1}{4\epsilon}.
\label{eq:app3}
\end{aligned}
\end{equation}
The density can be described by the power-law  distribution at very large radii, the above formula  can give its power index approximately(see the next paragraph in detail).
The above approximate analysis can help us to understand the physical process of dark matter halo from small scale to large scale.

\begin{figure}[H]
  \centering
  \includegraphics[angle=0,height=8.0cm,width=10.0cm]{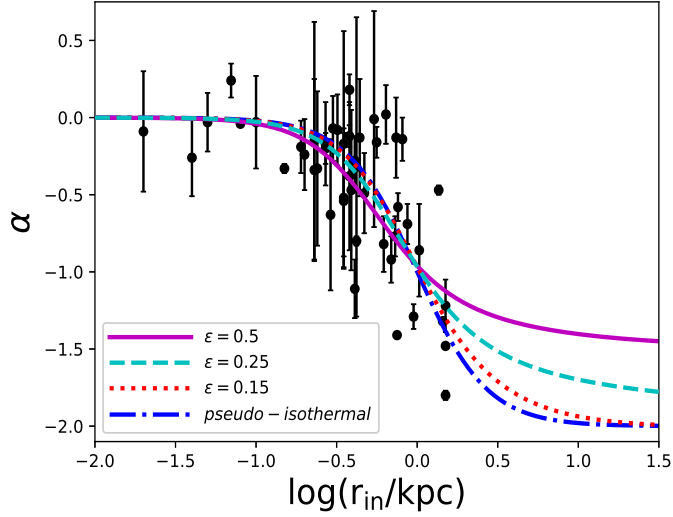}\\
    \caption{\small  Comparison results between the numerical profile in Equation (\ref{eq:gravity_bbb1}) and  the pseudo-isothermal profile. The solid line,  the dashed  line and the dotted line represent the numerical model with $\zeta=10^{-6}$ and $\epsilon=0.5,~ 0.25,~ 0.15$ respectively. The dash-dotted lines is for the pseudo-isothermal model with core radius of 1.0 kpc. Filled circles are the observed data of the slope $\alpha$, which are from the de Blok et al. (2001)\citep{2001ApJ...552L..23D} sample.
     }
  \label{Fig:index2}
\end{figure}

Because it is hard to get the analytical solutions of Equation(\ref{eq:gravity_aa1}) and Equation(\ref{eq:gravity_bbb1}),
  the  numerical solutions are necessary. The initial condition is $F=-(\frac{r}{r_{0}})^{2}$.
 Then the power indexes $\alpha$ for the numerical solutions  are showed in Fig. \ref{Fig:index2}.
  The observed data are also showed. The observed LSB sample involves 48 galaxies, which are
from de Blok et al. (2001)\citep{2001ApJ...552L..23D}. These numerical models can perfect response the observed result. The pseudo-isothermal halo model is also showed by the dash-dotted line in Fig. \ref{Fig:index2}. The numerical solution with $\epsilon=0.15$ is nearly same with the pseudo-isothermal halo model.  In Fig. \ref{Fig:index2}, it is easy to find that the values of $\alpha$ become flat at large radii, i.e. the density can be approximately described by the power-law  distribution. The rotation velocity $V_{rot}$  increases with radius, then $|F| \gg \zeta  $ at very large radii when $\epsilon \geq\frac{1}{4}$ and $\zeta>0$. The Equation (\ref{eq:app3})  can give the power index $\alpha$ approximately.

\begin{figure}[H]
  \centering
  \includegraphics[angle=0,height=8.0cm,width=10.0cm]{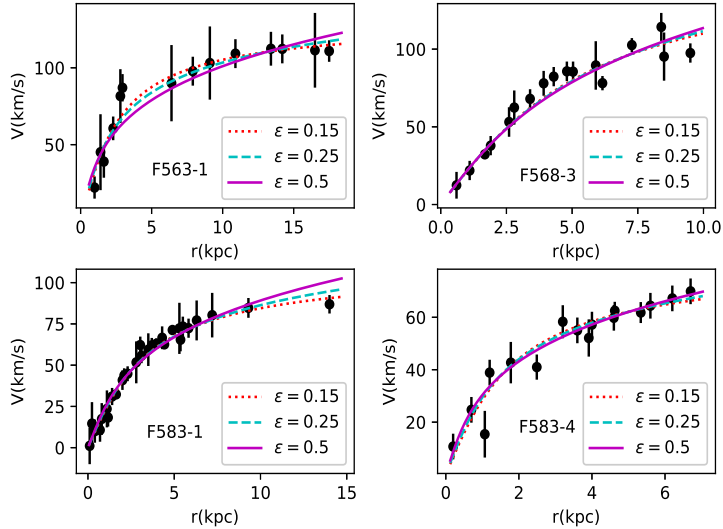}\\
    \caption{ \small  Observed LSB galaxy rotation curves with the best-fitting dark matter model. The solid line,  the dashed  line and the dotted line represent the numerical model with $\epsilon=0.5,~ 0.25,~ 0.15$, respectively. Filled circles stand for the observed data.
     }
  \label{Fig:vrot2}
\end{figure}

 The numerical solutions with $\epsilon=0.5,~ 0.25,~ 0.15$ are used in fitting the observed rotation curves. $\zeta$ and $r_{0}$ are adjustable parameters. The observed rotation curves of  LSB galaxies are from Kuzio de Naray et al. (2006, 2008, 2010)\citep{2006ApJS..165..461K, 2008ApJ...676..920K,2010ApJ...710L.161K}. The data are fitted by using the least-squares method. The fitting results are showed in Fig. \ref{Fig:vrot2}. The best fit parameter $\zeta$, $r_{0}$ and the reduced chi-square value $\chi_{\nu}^{2}$ are listed in Table 1, and $\zeta$ is expressed in SI units.

If $\epsilon=0$, because $|F|$ increases with radius as showed in the initial condition, and the term  $F^{2}$ reduces it, hence $F_{x}^{\prime}=0$ at $r=\infty$\citep{2010PhLB..694...10R}. Finally, this condition leads to $F\approx-4\zeta$ at $r=\infty$, i.e. the bigger $\zeta$ lead to the bigger rotate velocity. The rotation velocity $V_{rot}$ is equal to zero at galaxy center so that  $\zeta$ is proportional to the square of velocity dispersion at the galaxy center, the LSB galaxies with bigger velocity dispersion at the galaxy center should have bigger the peak rotation speed  ($V_{max}$). This phenomenon has been reported\citep{2006MNRAS.373..700B}.
The velocity dispersion at galaxy center is far less than the speed of light in Table 1. It indicates dark matter is cold\citep{2010PhLB..694...10R,2016PhLB..753..140P}.
Because the bigger velocity dispersion  can lead to the bigger  peak rotation speed, if we want to  $\rho$ drops more rapidly than the pseudo-isothermal density profile at the very outmost region, the term  $\zeta \rho$ must  vanish or the velocity dispersion must become smaller. In Fig. \ref{Fig:index2}, it shows that $\alpha$ can't be less than $-2$ at large scale. To let $\alpha$ is less than $-2$, the term  $\zeta \rho$ must  vanish at large scale, so we use the polytropic model to replace it in the following subsection.

\begin{table}[H]
\centering
  \noindent \small Table 1. The Best-fit parameters. \\
  [2mm]
   \begin{scriptsize}{
  \begin{tabular}{lccccccccc}
  \hline \hline
Galaxy & $\sqrt{\zeta/2}$ &$r_{0}$& $\chi_{\nu}^{2}$ & $\sqrt{\zeta/2}$ &$r_{0}$& $\chi_{\nu}^{2}$ & $\sqrt{\zeta/2}$ &$r_{0}$& $\chi_{\nu}^{2}$ \\
       & (km/s)           &  (kpc)&                  & (km/s)           &  (kpc)&                  &           (km/s) &(kpc)  &     \\
\hline
       & \multicolumn{3}{c}{$\epsilon=0.15$}  &\multicolumn{3}{c}{$\epsilon=0.25$}& \multicolumn{3}{c}{$\epsilon=0.5$} \\
\cmidrule(l){2-4} \cmidrule(l){5-7} \cmidrule(l){8-10}
F563-1 & 41.1$\pm$ 1.7 & 1.21$\pm$0.14 & 0.54 & 37.5$\pm$ 2.5 & 1.01$\pm$0.17 &0.72 & 31.4$\pm$ 5.4 &0.69$\pm$0.25& 1.08 \\
F568-3 & 52.1$\pm$ 4.4 & 2.49$\pm$0.26 & 1.28 & 54.4$\pm$ 5.3 & 2.58$\pm$0.31 &1.35 & 60.5$\pm$ 7.4 &2.83$\pm$0.42& 1.47 \\
F583-1 & 34.7$\pm$ 1.3 & 1.53$\pm$0.09 & 0.48 & 34.5$\pm$ 1.6 & 1.44$\pm$0.10 &0.57 & 34.8$\pm$ 2.7 &1.33$\pm$0.15& 0.80 \\
F583-4 & 26.0$\pm$ 1.4 & 0.81$\pm$0.11 & 0.67 & 24.5$\pm$ 1.7 & 0.68$\pm$0.11 &0.59 & 21.8$\pm$ 2.6 &0.49$\pm$0.12& 0.52 \\
   \hline \\
  \end{tabular}
  }
  \end{scriptsize}
\end{table}

\subsection{Case \uppercase\expandafter{\romannumeral3}}
\label{sect:model33}
\begin{figure}[H]
  \centering
  \includegraphics[angle=0,height=8.0cm,width=10.0cm]{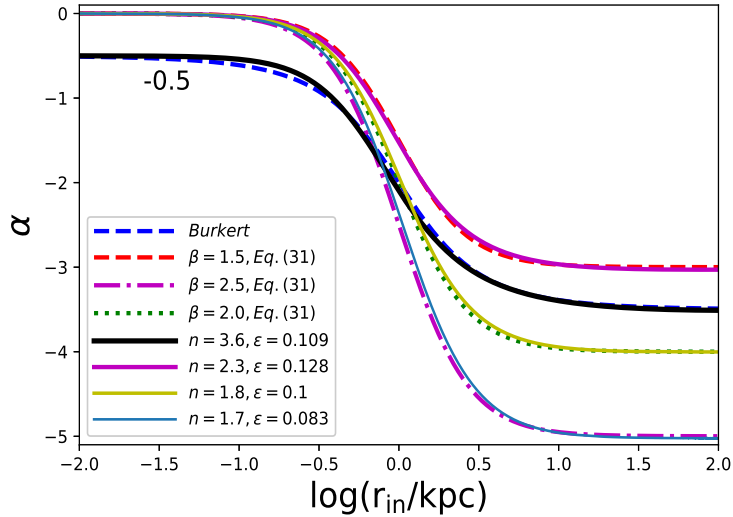}\\
    \caption{\small   The index $\alpha$ of numerical profile in Equation (\ref{eq:3gravity_bbb3}), the profile in Equation (\ref{eq:3gravity_profile}) and  the Burkert profile.  The solid lines from upper right to  bottom right represent the numerical model with $\zeta=10^{-7}$ and $(n,~\epsilon)=(2.3,~0.128),~ (3.6, ~0.109),~(1.8,~0.1),~(1.7,~0.083)$ respectively.
     The  dashed line at top, the dotted line, the dash-dotted line is for the profile in Equation (\ref{eq:3gravity_profile})
     with core radius of 1.0 kpc and $\beta=1.5,~2.0,~2.5$ (from top to bottom). The   dashed line at bottom represent the index $\alpha$ of Burkert profile.  To show clearly,  the index $\alpha$ of Burkert profile and the numerical model with $(n,~\epsilon)=(3.6,~ 0.109)$ minus 0.5.
     }
  \label{Fig:index33}
\end{figure}

There exist other density profiles which usually used in  LSB galaxies, such as thermal WDM halo density profile\citep{2016IJMPA..3150073D, 2010ApJ...710L.161K}
\begin{equation}
\begin{aligned}
\rho=\frac{\rho_{0}}{[1+(r/r_{0})^{2}]^{\beta}} ,~~~ 1 <  \beta \leq \frac{5}{2}
\label{eq:3gravity_profile}
\end{aligned}
\end{equation}
and the Burkert  density profile\citep{1995ApJ...447L..25B}
\begin{equation}
\begin{aligned}
\rho=\frac{\rho_{0}r_{0}^{3}}{(r^{2}+r_{0}^{2})(r+r_{0})}.
\end{aligned}
\end{equation}
Their indexes $\alpha$ are smaller than -2 in the very outer region.
To find similar solutions like above profiles, we assume that
\begin{equation}
\begin{aligned}
p= \frac{\zeta}{\rho_{0}^{s}}\rho^{1+s}+ 2\epsilon V_{rot}^{2}\rho,
\label{eq:3gravity_bbb3}
\end{aligned}
\end{equation}
where $s=\frac{1}{n}$ and $n$ is the polytropic index. This EOS includes the polytropic model.
Using the initial condition $F=-(\frac{r}{r_{0}})^{2}$, we get the numerical solutions of Equation (\ref{eq:gravity_aa1}) Equation (\ref{eq:3gravity_bbb3}).
Then the slopes $\alpha$ of the numerical solutions   are drawn in  Fig. \ref{Fig:index33}. To show clearly, the index $\alpha$ of Burkert profile  minus 0.5 in  Fig. \ref{Fig:index33}. When $n=5$ and $\epsilon=0$, the polytropic model can get the profile in  Equation (\ref{eq:3gravity_profile}) with $\beta=2.5$ in non-relativistic approximation, and it is nearly same  with the numerical solution with $(n,~\epsilon)=(1.7,~ 0.083)$ as showed in Fig. \ref{Fig:index33}. So there is degeneracy between  $n$ and $\epsilon$.

\section{Conclusions}
\label{sect:conclusions}

In this paper, we study  EOS of dark matter which is treated as a perfect fluid. When EOS is independent of the scaling transformation,
it does not explicitly contain x. Because the rotation velocity is far less than the speed of light, i.e. $F$ is very small,
we use the Taylor expansion to represent EOS approximately. Its first order terms can get that the pressure is proportional to density.
Its second order terms can  naturally  yield that the random motions of dark matter are correlated to the particle rotational motions.
Finally, we get a simplest EOS, i.e. $p=\zeta\rho+2\epsilon V_{rot}^{2}\rho$. It is not scale dependent. It can ensure a black hole and a constant-density core  hold simultaneously in one dark matter halo.

The term $\zeta\rho$ can lead to a constant-density core. The constant-density central core can exist in the region with $|F| \ll \zeta$.
The second order terms in the Taylor expansion  can lead to there exists a transition zone between  density core and outer region for power index $\alpha$ when density is characterized by power law relation. It can obtain a density profile which is similar to the pseudo-isothermal halo model when $\epsilon$ is around $0.15$. By means of the classical least chi square methodology, this profile can perfectly fit the observed rotation curves of LSB galaxies.

When $\zeta=0$, the term $\epsilon V_{rot}^{2}\rho$ can get a power law density beyond the region  that includes a black hole and of which mass is $\sqrt{2}$ times the black hole mass. The power index $\alpha$ is equal to $-\frac{1+4\epsilon}{4\epsilon}$.
When $\zeta > 0$ and $\epsilon$ is big, this power law density with $\alpha=-\frac{4\epsilon+1}{4\epsilon}$ also exist in the very outer region.
If $\zeta$ is proportional to the square of velocity dispersion at galaxy center, then the LSB galaxies with bigger velocity dispersion at the galaxy center should have bigger the peak rotation speed.

In order to  the constant random motions  vanish at large radii, we introduce the polytropic model. The polytropic model is scale dependent. For the equation of state that includes the polytropic model, i.e. $p= \frac{\zeta}{\rho_{0}^{s}}\rho^{1+s}+ 2\epsilon V_{rot}^{2}\rho$, we can get the density profiles with constant-density core and index $\alpha$  is less than -2 at  very large radii,  such as the profile which is nearly same with the Burkert profile. The polytropic model is widely used and can be obtained from many fundamental dark matter particle models, such as  Bose-Einstein condensate dark matter model\cite{2020EPJC...80..735C,2011PhRvD..84d3532C,2018PhRvD..98b3513D}, so  it is hard to be in favor of a dark matter particle model. The observations show that LSB galaxies nearly have a constant core column density, i.e. $\Sigma_{DM}=\rho_{0} r_{0}\approx 75 M_{\odot}pc^{-2}$\cite{2015ApJ...808..158B,2020ApJ...904..161B}. It may provide strong constraints on the physical properties of dark matter particle or core formation mechanism\cite{2020arXiv200704119M}. For example, the fuzzy dark matter model  is hard to explain it\cite{2020ApJ...904..161B}. $\rho_{0}$ and $r_{0}$ in our model are free parameters, and their relationship is not involved in this paper.  We can't explain the observation about constant core column density now. We will focus on it in our future work.

~\\

\acknowledgments{This work is  supported by the National Natural Science Foundation
(NSF) of China (No. 11973081, 11573062, 11403092, 11390374, 11521303),
the YIPACAS Foundation (No. 2012048), the Chinese Academy of Sciences
(CAS, KJZD-EW-M06-01), the NSF of Yunnan Province (No. 2019FB006) and
the Youth Project of Western Light of CAS.

Software: NumPY \citep{5725236}, SciPy \citep{2020NatMe..17..261V}, Matplotlib \citep{Hunter:2007}.}

%\end{multicols}

%\vspace{10mm}

\vspace{-1mm}
\centerline{\rule{80mm}{0.1pt}}
\vspace{2mm}

%\begin{multicols}{2}
\bibliographystyle{unsrt}

%\end{multicols}

\clearpage
\end{CJK*}
\end{document}